\documentclass[letterpaper,twocolumn]{jpsj3}
\usepackage[font=it]{caption}
\usepackage{lipsum,graphicx,float}
\usepackage{authblk}
\usepackage[hyphens]{url}
\usepackage{txfonts}
\usepackage[bottom]{footmisc}
\usepackage[shortcuts]{extdash}

\title{Quanvolutional Neural Networks: Powering \\Image Recognition with Quantum Circuits}
\author{Maxwell Henderson$^1$\thanks{max.henderson@qxbranch.com}, Samriddhi Shakya$^1$, Shashindra Pradhan$^1$, and Tristan Cook$^1$}
\inst{$^1$QxBranch, Inc., 777 6th St NW, 11th Floor, Washington DC, 20001}

\abst{Convolutional neural networks (CNNs) have rapidly risen in popularity for many machine learning applications, particularly in the field of image recognition.  Much of the benefit generated from these networks comes from their ability to extract features from the data in a hierarchical manner.  These features are extracted using various transformational layers, notably the convolutional layer which gives the model its name.  In this work, we introduce a new type of transformational layer called a quantum convolution, or quanvolutional layer.  Quanvolutional layers operate on input data by locally transforming the data using a number of random quantum circuits, in a way that is  similar to the transformations performed by random convolutional filter layers.  Provided these quantum transformations produce meaningful features for classification purposes, then the overall algorithm could be quite useful for near term quantum computing, because it requires small quantum circuits with little to no error correction.  In this work, we empirically evaluated the potential benefit of these quantum transformations by comparing three types of models built on the MNIST dataset: CNNs, quantum convolutional neural networks (QNNs), and CNNs with additional non\=/linearities introduced.  Our results showed that the QNN models had both higher test set accuracy as well as faster training compared to the purely classical CNNs.}

\begin{document}
\maketitle
\section{Introduction} \label{introduction}

The field of quantum machine learning (QML) has experienced rapid growth over the past few years, as evidenced by the rapid increase in impactful QML papers \cite{quantum_zoo}.  Several excellent papers \cite{dunjko_2018, ciliberto2018quantum, dunjko_2016, alejandro_qml, biamonte2017quantum} encapsulate the current state of the QML field.  Machine learning algorithms tend to give probabilistic results and contain correlated components but at the same time suffer computational bottlenecks due to the curse of dimensionality.  Similarly, quantum computers by their very nature provide probabilistic results upon measurement and are formed from intrinsically coupled quantum systems, which can provide potentially exponential speedups due to their ability to perform massively parallel computations on the superposition of quantum states.  While quantum computers are by no means expected to replace classical computing, they have have the potential to be powerful components in an overall machine learning application pipeline.

Our research focuses on a novel quantum algorithm which falls squarely into the regime of ``hybrid classical-quantum" algorithms, extending the classical algorithm of convolutional neural networks (CNNs).  In the few short years since their inception in by LeCun \cite{cnn}, CNNs have become the standard for many machine learning applications. Although the emerging capsule networks \cite{capsule_networks} have shown potential promise for pushing the bounds of machine learning performance even further, various flavors of CNNs have held the accuracy records on many benchmark image recognition problems for years, including MNIST, CIFAR, and SVHN \cite{machine_learning_benchmarks}.

CNNs can be visualized as a ``stack" of transformations that are applied to input data.  These transformations are applied in an attempt to extract useful features in the data which can be leveraged for classification purposes.  The convolutional layers in a CNN stack are each composed of $n$ convolutional filters.  Each filter in a convolutional layer iteratively convolves local subsections of the full input to produce feature maps, and the output of the convolutional layer will be a tensor of $n$ feature maps which each contain information about different spatially-local patterns in the data.  Each layer of the CNN repeats this process on the output of the layer preceding it, resulting in increasingly abstract features.  These abstract features are useful because classifiers built on top of these transformations (or more likely a series of transformations through an entire network stack) produce far more accurate results than classifiers built directly on top of the input data itself.

In this work, we investigate a new type of model which we will call \textit{quanvolutional} neural networks (QNNs).  QNNs extend the capabilities of CNNs by leveraging certain potentially powerful aspects of quantum computation.  QNNs add a new type of transformational layer to the standard CNN architecture: the quantum convolutional (or \textit{quanvolutional}) layer.  Quanvolutional layers are made up of a group of $N$ quantum filters, which operate much like their classical convolutional layer counterparts by producing features maps through locally transforming input data.  The key difference is that quanvolutional filters extract features from input data by transforming spatially-local subsections of data using random quantum circuits.  We hypothesize that features produced by random quanvolutional circuits could increase the accuracy of machine learning models for classification purposes.  If this hypothesis holds true, then QNNs could prove a powerful application for near term quantum computers, which are often called noisy (not error corrected), intermediate-scale (50 - 100 qubits) quantum (NISQ) computers \cite{preskill_nisq}.  This is primarily due to two reasons: (1) quanvolutional filters are applied to only local subsections of input data, so they can operate using a small number of quantum bits (qubits) with shallow gate depths and (2) quanvolutions are resilient to error; as long as the error model in the quantum circuit is consistent, it can essentially be thought of as yet another component in the random quantum circuit. Finally, since determining output of random quantum circuits is not possible to simulate classically at scale, if the features produced by quanvolutional layers provide a modeling advantage, they would require quantum devices for efficient computation.

\section{Architectural design of QNNs } \label{qnn_architecture}

\subsection{Design motivations}
This section serves as a motivational preface to the use of quanvolutional transformations within a broader machine learning framework.  First, leveraging random nonlinear features is well-known to be useful within many machine learning algorithms for increasing accuracy or decreasing training times, such as in CNNs using random convolutions and echo state networks \cite{Jaeger2004, Ranzato2007}.  Second, as pointed out in Mitarai etl al. 2018 \cite{quantum_circuit_learning}, quantum circuits are able to model complex functional relationships, such as universal quantum cellular automata, which is infeasible using polynomial-sized classical computational resources.  The idea of combining both of these observations together - leveraging some form of non-linear quantum circuit transformations for machine learning purposes - has recently emerged in the QML field.  We may consider briefly the current state of the QML field with algorithms in this space, as well as the potential ``quantum advantage" motivations for such algorithms.

Several quantum variations of classical models have been recently developed, including quantum reservoir computing (QRC) \cite{Fujii2016}, quantum circuit learning (QCL) \cite{quantum_circuit_learning}, continuous-variable quantum neural networks \cite{Killoran2018}, quantum kitchen sinks (QKS) \cite{quantum_kitchen_sinks}, quantum
variational classifiers and quantum kernel estimators \cite{Havlicek2019}.  The QNN approach similarly aims to use the novelty of quantum circuit transformations within a machine learning framework, while differing from previous works in (a) the particular methodology around processing classical information into and out of the different quantum circuits (more details in \ref{quantum_filter_design}) and (b) the flexible integration of such computations into state\=/of\=/the\=/art deep neural network machine learning model.  Another key difference in the QNN approach revolves around the lack of variational tuning of the quantum components of the model.  Several different frameworks have been presented for the use of variational methods for training the quantum components of such quantum feature transformation methods \cite{Bergholm2018, Crooks2018, Schuld2018}.  This work however builds off of the static feature methodology, in some sense extending the work of the QRC methodology.

In terms of a quantum advantage or quantum supremacy consideration of this model, this paper makes an argument similar to that of Havlicek et al. \cite{Havlicek2019}. In essence, QNNs can efficiently access kernel functions in high dimensional Hilbert spaces, which if useful for machine learning purposes, could provide a pathway to quantum advantage.

\subsection{Quanvolutional network design} \label{quanvolutional_network_design}
As laid out briefly in Section \ref{introduction}, QNNs are simply an extension of classical CNNs, with an additional transformational layer, called the quanvolutional layer.  Quanvolutional layers should integrate exactly the same as classical convolutional layers, which allow the user to:
\begin{itemize}
\item Define any arbitrary integer value for the number of quanvolutional filters in a particular quanvolutional layer
\item Stack any number of new quanvolutional layers on top of any other layer in the network stack
\item Provide layer-specific configurational attributes (encoding and decoding methods, average number of quantum gates per qubit in the quantum circuit, etc)
\end{itemize}

\begin{figure*}[h]
    \centering
    \includegraphics[width=1\textwidth]{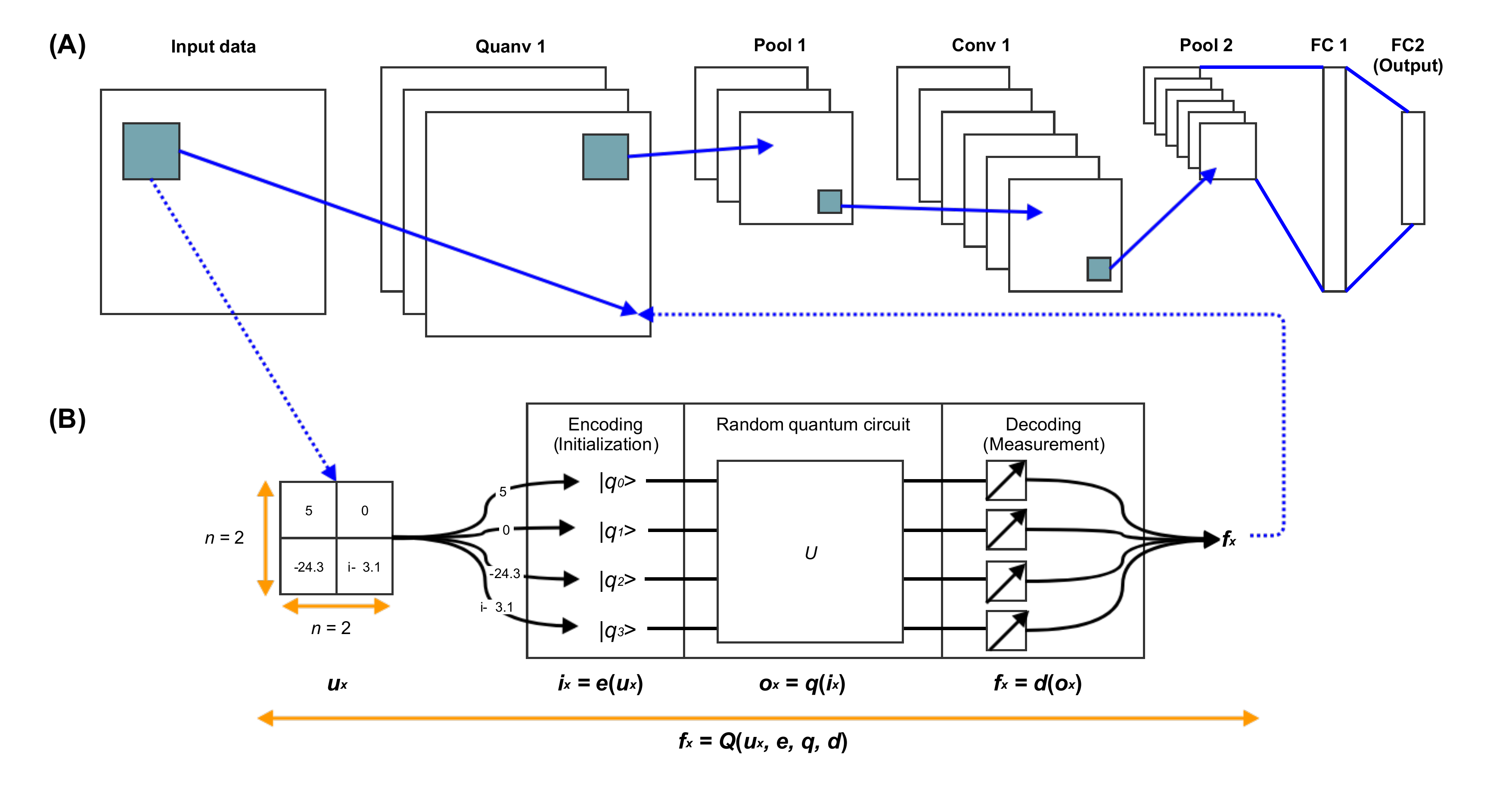}
  \caption{A. Simple example of a quanvolutional layer in a full network stack.  The quanvolutional layer contains several quanvolutional filters (three in this example) that transform the input data into different output feature maps.  B. An in-depth look at the processing of classical data into and out of the random quantum circuit in the quanvolutional filter.}
  \label{fig:qnn_and_filter_architecture}
\end{figure*}

Satisfying these conditions, the quanvolutional filter then should be very generalizable and just as easy to implement in any architecture as its classical predecessor.  The number of such layers, the order in which they are implemented, and the particular parameters of each are left entirely up to the end user's specifications.  The generality of QNNs is visualized in Fig. \ref{fig:qnn_and_filter_architecture}.  Figure \ref{fig:qnn_and_filter_architecture}.A shows just one example QNN realization, where the first layer of the stack has a quanvolutional layer with 3 quanvolutional filters, followed by a pooling layer, a convolutional layer with 6 filters, a second pooling layer, and two final fully\=/connected (FC) layers, wherein the final fully-connected layer represents the target variable output.  The diagram conveys the generality and flexibility for architects to change, remove, or add layers as desired.  The overall network architecture of Fig. \ref{fig:qnn_and_filter_architecture}.A would have the exact same structure if we replaced the quanvolutional layer with a convolutional layer of three filters, or similarly if we replaced the convolution layer with a quanvolutional layer with six quanvolutional filters.  The difference between the quanvolutional and convolutional layer depends on the way that the quanvolutional filters of Fig. \ref{fig:qnn_and_filter_architecture}.B perform calculations, which is laid out in detail in Section \ref{quantum_filter_design}.

\subsection{Quanvolutional filter design} \label{quantum_filter_design}

Quanvolutional filters each produce a feature map when applied to an input tensor, by transforming spatially-local subsections of the input tensor using the quanvolutional filter.  However unlike the simple element-wise matrix multiplication operation that a classical convolutional filter applies, a quanvolutional filters transforms input data using a quantum circuit, which can be structured or random.  For simplicity and to establish a baseline, in this work we use random quantum circuits for the quanvolutional filters as opposed to circuits with a particular structure.

At a high level, quanvotutional filters transform input and output a scalar calculated from a 2D matrix of scalars by a universal quantum computing (UQC) circuit in the set BQP \cite{bqp}.  We can formalize this process for transforming classical data using quanvolutional filters as follows:
\begin{enumerate}
\item Let us consider a single quanvolutional filter.  This quanvolutional filter uses a random quantum circuit $q$, which takes as input spatially-local subsections of images from dataset $u$.  For our purposes, we define each of these inputs as $u_{x}$, and each $u_x$ will be a 2D matrix of size $n$-by-$n$ wherein $n>1$.
\item Although there are many ways of ``encoding" $u_{x}$ as an initialized state of $q$, for each quanvolutional filter we choose one particular encoding function $e$; we define the encoded initialization state $i_{x} $ as $i_{x} = e(u_{x})$.
\item After the quantum circuit is applied to the initialized state $i_{x}$, the result of the quantum computation will be an output quantum state $o_{x}$, with the relationship $o_{x} = q(i_{x}) = q(e(u_{x}))$.
\item Although there are many ways of ``decoding" the information made available about $o_{x}$ through a finite number of measurements, to ensure that the quanvolutional filter output is consistent with similar output from a standard classical convolution, we define the final decoded state as $f_{x} = d(o_{x}) = d(q(e(u_{x})))$ wherein $d$ is our decoding function and $f_{x}$ is a scalar value.
\item Let us define the total transformation of $d(q(e(u_{x})))$ from this point on as the ``quanvolutional filter transformation" $Q$ of $u_{x}$, aka $f_{x} = Q(u_{x}, e, q, d)$.  In Figure \ref{fig:qnn_and_filter_architecture}.B a visualization of a single quanvolutional filter is shown, displaying the encoding / applied circuit / decoding process.
\item If we consider the number of calculations that occur when applying a classical convolution filter to input from dataset $u$, it is clear that the number of computations required is simply $\mathcal{O}(n^{2})$, placing the computational complexity squarely in P.  This is not the case for the computational complexity of $Q$, which is \#P-hard \cite{huang_quantum_simulation_complexity}; this emerges specifically out of the complexity of the random quantum circuit transformation $q$, while $e$ and $d$ can be performed efficiently on classical devices.
\end{enumerate}

Our experimental argument is the following: let us consider a theoretical situation in which an end user has access to many quanvolutional filters within a quanvolutional layer, with the goal of building a model on dataset $u$ for classification purposes (i.e., generate a model that can properly label an input image with a correct output label).  In this example, suppose that we build two types of networks: (1) networks built using solely classical convolutional layers and (2) networks using at least one quanvolutional layer.  If the networks of type 1 result in better performance results compared to type 2, a reasonable insight to draw from this outcome is that the quantum features extracted from the training dataset $u$ were not ultimately useful in building abstract features for classification purposes.  However, if networks of type 2 routinely and significantly outperformed networks of type 1, we could again reasonably draw an insight that the quantum features produced were useful in building features on dataset $u$ for classification purposes.  Additionally, this could signal a true ``quantum" advantage, as the classical generation of the quantum features, as described earlier, would fall squarely within BQP.

\subsection{Strengths and limitations of quanvolutional approaches} \label{caveats}
While the potential benefits of quanvolutional filters were expanded on in Section \ref{quantum_filter_design} compared to classical, there are several benefits compared to many other quantum computing algorithms that make the QNN approach ideal for the NISQ computing era.
\begin{itemize}
\item \textit{Inherently hybrid algorithm}.  All NISQ algorithms of interest will be hybrid by nature, and the QNN framework embraces this whole-heartedly.  The QNN stack purposely integrates elements from both classical data and algorithms (CNNs) with quantum subprocesses (quanvolutional layers).  While inherently hybrid classical-quantum approaches will likely fall short of some of the lofty goals of quantum computing, such as exponential speedups over classical calculations, we contend that the NISQ field will show a ``crawl-walk-run" evolution, where modest (polynomial) speedups and improvements will occur before larger (exponential) ones are discovered.
\item \textit{No QRAM requirements}.  As pointed out in Aaronson 2015 \cite{Aaronson2015}, a major hurdle for QML speedups is the lack of an efficient way of loading large classical data into quantum random access memory (QRAM) for future operations.  However, since the QNN framework approach simply requires running \textit{many} quantum circuits on \textit{single} data points, there is no need to store entire datasets into QRAM. 
\item \textit{Potential resiliency to unknown but consistent error models}.  Since quanvolutional layer feature maps result from transforming data through random quantum circuits, it is reasonable to assume that adding in particular error models does not necessarily invalidate the overall algorithm.  Conceptually, many forms of quantum error can be thought of as unknown, and unwanted, gate operations.  For example, the user attempts to run a particular quantum circuit, and due to hardware imperfections and limitations, some subtle and hidden ``noisy" quantum gates are also added into the circuit, resulting in a final quantum state different than desired.  Since QNNs use the quantum circuits as feature detectors, and it is not clear a priori which quantum circuits leads to the most useful features, adding in some unknown ``noise" gates does not necessarily impact feature detection quality overall.  We consider that modeling non-stable, noisy error effects of NISQ devices in quanvolutional filters is an open question, hence assessing how well the QNN approach performs in the presence of such error is outside the scope of this paper.
\end{itemize}

While these points are encouraging of NISQ-era viability, one must also be aware of the limitations, constraints, and open questions of the overall QNN approach.
\begin{itemize}
\item \textit{Optimal interface with classical data}.  As described in Section \ref{quantum_filter_design}, two key components of the quantum filter design are the encoding and decoding of the classical information and how they interface with the quantum system.  While the particular encoding and decoding protocols used in this experiment will be discussed in more detail in Section \ref{quantum_filter_details}, it remains an open question about how to optimally design these protocols.  Further, even if some protocols were determined to be useful, some may be impractical in experimental use.  For instance, decoding methods that require more than one measurement of the quantum system quickly become problematic; referencing again Aaronson 2015 \cite{Aaronson2015}, any potential ``quantum speedup" disappears for algorithms that require large numbers of quantum measurements.  Ideally then, encoding and decoding methods are required which to (1) produce beneficial machine learning features and (2) require minimal measurements.
\item \textit{Large number of transformations}.  Although the quanvolutional filter designs have the advantage that QRAM is not required, they may require a large number of quantum circuit executions.  This becomes readily apparent if one considers the number of  computations that are required in a normal convolutional layer.  As an example, the number of computations for a convolutional layer with 50 convolutional filters using zero-padding and a stride of 1, on an input 100\=/by\=/100\=/by\=/1 pixel image, then $50\times100\times100=500,000$ element-wise matrix multiplication operations will need to be applied on this image alone.  This presents challenges for using quanvolutional layers; if using the same number of quanvolutional filters on the same input data we would require the same number of quanvolutional filter executions.  While performing large numbers of quantum computations may not present as much of a strain as the hardware becomes more prevalant and mature, in the NISQ-era this would be impractical in general.  Strategies to work around this limitation are outlined in Section \ref{quantum_filter_details}.
\item \textit{Clear demonstration of quantum advantage}.  Experimentally verifying a near-term algorithm which shows quantum computing advantage over classical methods continues to be highly elusive.  While some groups continue to work towards particular experiments that could definitively show some form of ``quantum supremacy" \cite{Boixo2016}, the QNN algorithm laid out in this work aims at providing a general framework and strategy for future such endeavors.  However, ultimately the burden of proof is on the QNN algorithm to show why such an approach provides any benefit over other classical transformations.  Showing that the QNN approach can be useful in a machine learning context is a good starting point, but true adoption of such quantum methods should only come after a clear benefit is displayed.  These discussions and comparisons will be investigated in detail in Section \ref{results}.
\end{itemize}

\section{Experimental design} \label{methodology}
Our experiments in this work were designed to highlight the novelties introduced by the QNN algorithm: the generalizability of quanvolutional layers inside a typical CNN architecture, the ability to use this quantum algorithm on practical datasets, and the potential use of features introduced by the quanvolutional transformations. There has been an increase in research into the use of quantum circuits in machine learning applications lately, such as in Wilson et al. 2018 \cite{quantum_kitchen_sinks}, where classical data is processed through randomly parameterized quantum circuits and the output is then used to train linear models. These models built on the quantum transformations were shown to be beneficial compared to other linear models built directly on the dataset itself, but did not have the same level of performance compared to other classical models (such as SVMs). The experiments in this work build off these results by building this quantum feature detection into a more complex neural network architecture, as the QNN framework naturally introduces classical models that already contain non-linearities.  In Section \ref{tested_models}, we clearly specify the comparisons between quantum and classical approaches.
\subsection{Classical dataset} \label{classical_dataset}
The image benchmark MNIST dataset \cite{cnn} was used in this work, which contains 70,000  (60,000 training and 10,000 test) 28\=/by\=/28 greyscale pixel images.  While several QML papers used a subset of the full dataset \cite{quantum_kitchen_sinks} or reduced the dimensionality of the dataset to fit onto hardware \cite{Adachi2015, arbitrary_pairwise}, the spatially-local transformational nature of the quanvolutional approach allows this framework to be applied to large, high-dimensional datasets.
\subsection{Tested models} \label{tested_models}
In this research, we tested three separate models: 
\begin{enumerate}
\item \textit{CNN MODEL}.  A purely classical convolutional neural network, with the following network structure: CONV1 - POOL1 - CONV2 - POOL2 - FC1 - FC2.  Each convolutional layer used ReLU and filters of size 5-by-5, and the first and second convolutional layers had 50 and 64 filters, respectively.  The first fully-connected layer had 1024 hidden units and a dropout layer of 0.4, and as second fully-connected layer is the output layer, with 10 hidden units (1 for each target variable label).
\item \textit{QNN MODEL}.  The most basic quanvolutional neural network: a CNN network with a single quanvolutional layer.  Specifically, the single quanvolutional layer is the first transformation in the stack, and then the remaining architecture on top is the exact same as the CNN MODEL: QUANV1 - CONV1 - POOL1 - CONV2 - POOL2 - FC1 - FC2.  The number of filters in the quanvolutional layer swept over a range from 1 to 50 to determine the effect of the number of filters on model performance. 
\item \textit{RANDOM MODEL}.  A similar network to QNN MODEL architecture, except instead of the first transformation being a quanvolutional transformation, a purely classical random non\=/linear transformation was applied instead.
\end{enumerate}
By comparing the QNN MODEL network to both the CNN MODEL and the RANDOM MODEL, we can address whether or not adding quantum features improves in any way the overall CNN model performance, and investigate the QNN performance against a classical non-linear approach.  Each model was trained for 10,000 iterations and at each 100 training steps, the current log-loss and test set accuracy results of the model were saved.
\subsection{Experimental environment} \label{qcss}
The experiments performed in this research were conducted on the QxBranch Quantum Computer Simulation System. This experimental environment represents a universal quantum computer capable of executing gate\=/model instructions of abitrary gate width, circuit depth, and fidelity. No noise models were used in the experiments so as to assess the effectiveness of the ideal universal quantum computational model versus the classical computational model. Future experimentation could include the effects of noise, or use of NISQ hardware such as provided by Google, IBM and Rigetti Computing.
\subsection{Quanvolutional filter generation methodology}  \label{quantum_filter_generation}
To generate each quantum filter, we require the input size of the quanvolutional filters, which defines how many qubits are required for the circuit\footnote{It is not a hard requirement in general that the number of qubits in the quanvolutional filter matches the dimensionality of the input data.  In many quantum circuit applications, ancilla qubits are required to perform elements of an overall computation.  Such transformations using ancilla qubits, however, are outside the scope of this work and will be an interesting future topic to explore.}. In this work, we chose the simplest possible implementation to use only 3-by-3 quanvolutional filters, so that each simulated circuit had exactly 9 qubits ($n=3$).  We then generated the actual circuit by treating each qubit as a node in a graph, and assigning a ``connection probability" between each qubit.  This probability is the likelihood that a 2\=/qubit gate will be applied between the two qubits in question, using either a randomly selected CNot, Swap, SqrtSwarp, or ControlledU gate.  Additionally, a random number (in the range $[0, 2n^2]$) of 1\=/qubit gates (coming from the gate set $[X(\theta), Y(\theta), Z(\theta), U((\theta), P, T, H]$, where $\theta$ is a random rotational parameter) were generated, with the target qubit for each gate chosen at random.  After all 1 and 2\=/qubit gates are selected, the order of these gate operations in the circuit is shuffled.  This final ordering of gate operations becomes one quanvolutional filter in the quanvolutional layer.
\subsection{Encoding and decoding methodology} \label{quantum_filter_details}
As mentioned in Section \ref{caveats}, the optimal method for interfacing between the classical and quantum components of the QNN algorithm is currently an open question.  In our approach, the gate operations in our quantum circuits were kept static, but the information was encoded into the initial states of each individual qubit, as shown in Figure \ref{fig:qnn_and_filter_architecture}.B.  To experiment with the simplest form of encoding, we simply applied a threshold value to each pixel, and values greater than this threshold (in this case 0) were encoded in the $|1\rangle$ state, while those equal to or below were encoded in the $|0\rangle$ state.  In terms of decoding, as shown in Figure \ref{fig:qnn_and_filter_architecture}.B, to enforce that the quanvolutional network functions in a similar way to a convolutional filter, the output decoding is condensed to a scalar output value. Taking advantage of the QCSS' ability to output the whole state vector, we select the most likely output qubit vector state and sum the number of qubits that were measured in the $|1\rangle$ state.  This drastically reduces the total input state-space, and made it possible to, as a pre-processing step, fully determine all possible input-output mappings for any input data.  This means that in essence, a look-up table was being applied on each new input set of spatially-local data, rather than re-running the data through a quantum circuit each time. While this exact approach is infeasible on hardware since the full state-space will never be known, for this initial research we still choose to use this method to best understand what results would be expected the case of full-information.  In future work, a more practical, similar approach using a finite number of number of measurements could be implemented, taking the most commonly measured result.
 
\section{Results} \label{results}

\begin{figure}[H]
	\centering
	\includegraphics[width=\linewidth]{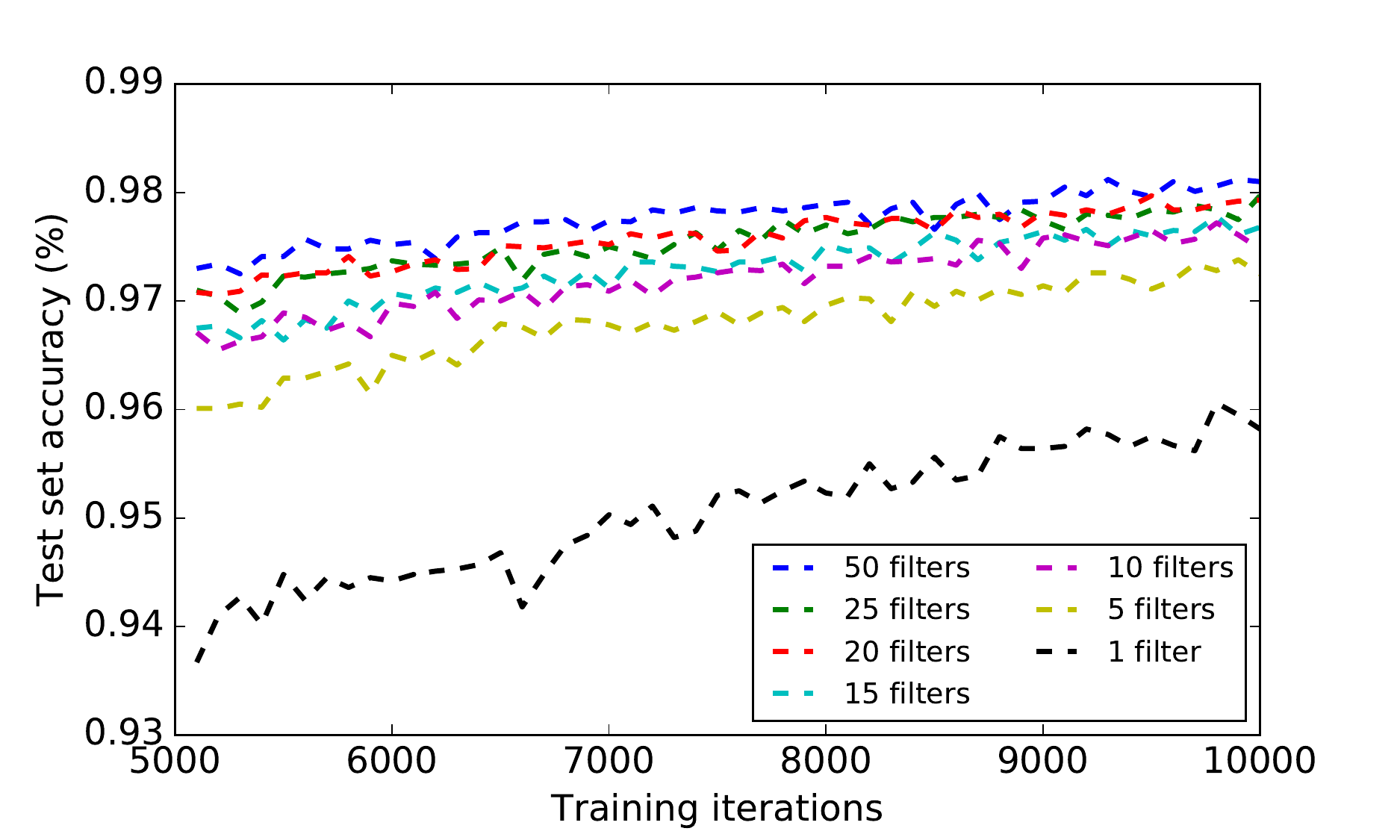}
  \caption{QNN MODEL test set accuracy results using a variable number of quanvolutional filters.}
  \label{fig:accuracy_filters}
\end{figure}

Before comparing the overall QNN algorithm to classical performance, it is imperative that we first validate that the overall algorithm is performing as expected.  To do this, we first run several different QNN MODELs with a varying number of quanvolutional filters and analyze the test set accuracy as a function of training iteration.  These results are shown in Figure \ref{fig:accuracy_filters}, and validate two important aspects of the overall QNN algorithm.  First, the QNN algorithm functioned as expected within the larger framework; adding the quanvolutional layer to the overall network stack generated the high accuracy results (95\% or higher) expected from a deep neural network.  Secondly, the quanvolutional layer and network seemed to behave as expected compared to using a similar classical convolutional layer.  The more training iterations, the higher the model accuracy became.  Additionally, adding more quanvolutional filters increased model performance, consistent with adding more classical filters to a convolutional network.  This observation reaches convergence ax expected; just as a standard convolutional layer reaches a ``saturation" effect after a certain number of filters, a similar convergence was observed for the number of quanvolutional filters.  While the network accuracy increased radically from a single filter to 5, and similarly from 5 to 10, there was minimal advantage in using 50 vs 25 quanvolutional filters in this experiment.

\begin{figure}[H]
	\centering
	\includegraphics[width=\linewidth]{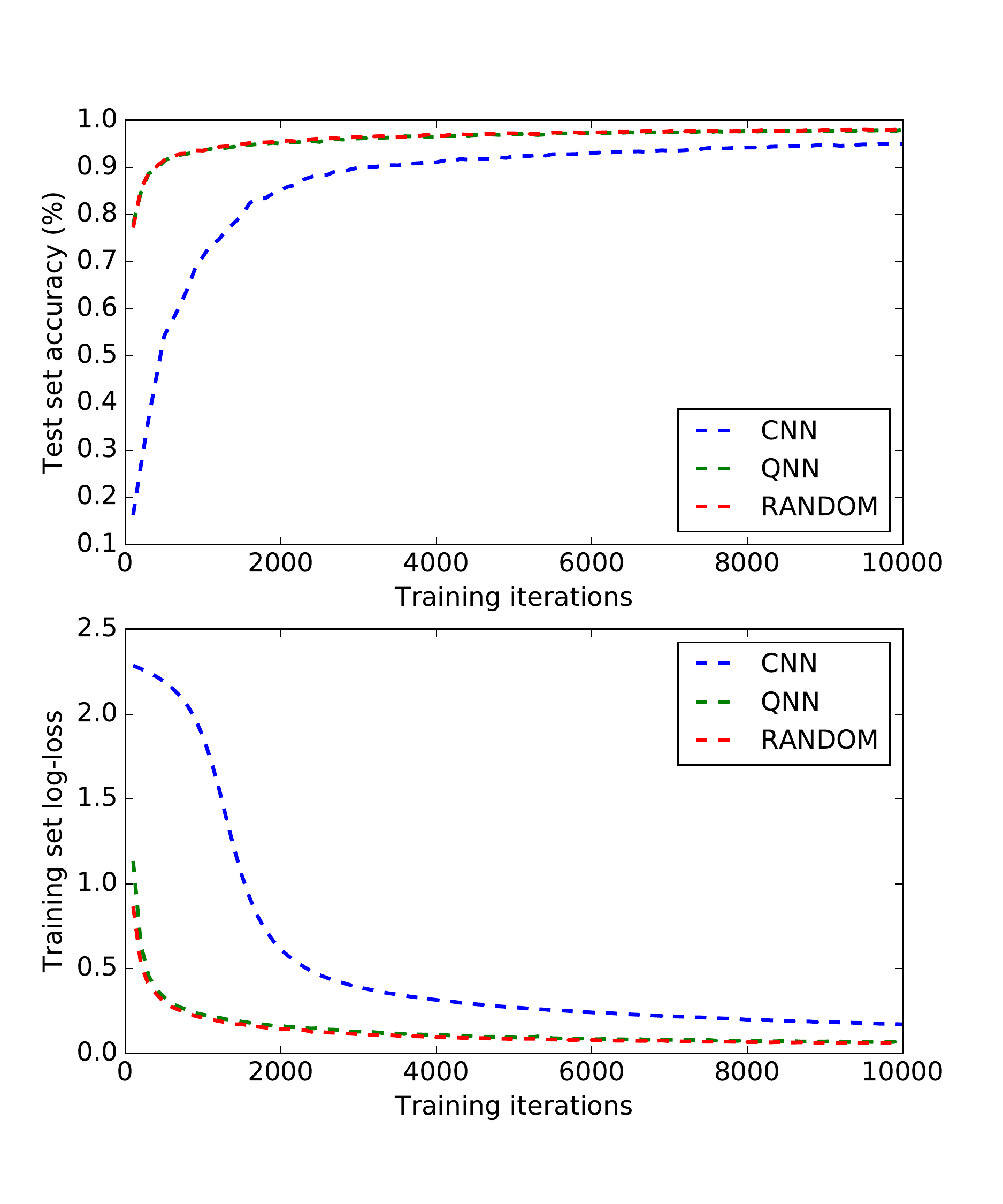}
  \caption{QNN MODEL performance, in terms of (A) test set accuracy and (B) training log-loss, compared to both CNN MODEL and RANDOM MODEL.}
  \label{fig:combined_accuracy_logloss}
\end{figure}

Having validated that the overall quanvolutional layer was performing as intended, the final test was to determine how these QNN MODELs compared to both CNN MODEL and RANDOM MODEL.  In our experiment, both the QNN MODEL and the RANDOM MODEL had 25 transformations in the quanvolutional and random, non-linear transformational layer, respectively.  The results comparing these three models are shown in Figure \ref{fig:combined_accuracy_logloss}.

The results of Figure \ref{fig:combined_accuracy_logloss} exhibit a consistency with those of Wilson et al. 2018 \cite{quantum_kitchen_sinks}, as well as showing some results that extend the scope of quantum features (features produced by by processing classical data through quantum transformations) applicability even further.  This work \cite{quantum_kitchen_sinks} showed that quantum transformations feeding into a linear model could give a performance enhancement over linear models built on the data directly.  In a similar way, the performance boost seen in Figure \ref{fig:combined_accuracy_logloss} makes a strong argument that adding in quantum features feeding into a more complex non-linear model stack can lead to a similar performance benefit compared to the same non-linear model stack built directly on the data itself.  However, while this is a promising step for some form of QNN applicability, it still does not show any clear quantum advantage over all classical models.  The results of RANDOM MODEL were statistically indistinguishable from the QNN MODEL results; this would imply that the non-linear transformations of the random quantum circuits were not in any significant way advantageous (or disadvantageous) compared to classical random non-linear transformations.

\section{Conclusions}

QNNs could provide early quantum adopters a useful, flexible, and scalable quantum machine learning application for real-world problems.  The QNN experimental results showed that even within a larger, typical deep neural network architecture stack, quanvolutional transformations to classical data can increase accuracy in the network; however, this research did not definitively show any quantum advantage over other classical, non-linear transformations.  Quanvolutional filter transformations may present some potential form of ``quantum advantage" if (1) such transformations prove useful for classification purposes on a particular dataset(s) and are (2) classically intractable to simulate at scale.  Although a massive number of practical questions need further research to bring the QNN functionality to its peak performance, if these questions can be answered then the overall framework presents a strong candidate for a success story for the NISQ era of quantum computing.

While outside the scope of this work, these results would seem to paint a clear path forward in terms of using a QNN framework: the true challenge will be to determine what properties of quanvolutional filters, or what filters in particular, are both (1) useful for machine learning purposes and (2) classically difficult to simulate.  This presents interesting challenges and research questions.  Are there particular structured quanvolutional filters that seem to always provide an advantage over others?  How data-dependent is the ideal ``set" of quanvolutional filters?  How much do encoding and decoding approaches influence overall performance?  What are the minimal quanvolutional filter gate depths that lead to some kind of advantage?  Attempting to find answers for these (and similar) questions will be necessary steps towards any path towards a viable use case of QNNs.

\section{Acknowledgements}
We would also like to thank Duncan Fletcher for your significant editing input, and helpful comments by Peter Wittek.

\bibliographystyle{jpsj3}
\bibliography{qnn}


\end{document}